\documentstyle[epsfig]{elsart}
\begin{document}
\begin{frontmatter}
\title{K$^+$ production in the reaction $^{58}$Ni+$^{58}$Ni at
incident energies from 1 to 2 AGeV}

\author[gsi]{D.~Best},
\author[gsi,hd]{N.~Herrmann},
\author[gsi]{B.~Hong},
\author[gsi,warsaw]{M.~Kirejczyk},
\author[gsi]{J.~Ritman},
\author[gsi,warsaw]{K.~Wisniewski},
\author[moscow]{A.~Zhilin},
\author[gsi]{A.~Gobbi},
\author[gsi]{K.~D.~Hildenbrand},
\author[gsi]{Y.~Leifels},
\author[gsi]{C.~Pinkenburg},
\author[gsi]{W.~Reisdorf},
\author[gsi]{D.~Sch\"ull},
\author[gsi]{U.~Sodan},
\author[gsi]{G.~S.~Wang},
\author[gsi]{T.~Wienold},
\author[clermont]{J.~P.~Alard},
\author[clermont]{V.~Amouroux},
\author[clermont]{N.~Bastid},
\author[moscow]{I.~Belyaev},
\author[hungary]{G.~Berek},
\author[rossendorf]{J.~Biegansky},
\author[kurchatov]{R.~Cherbatchev},
\author[strasbourg]{J.~P.~Coffin},
\author[strasbourg]{P.~Crochet},
\author[clermont]{P.~Dupieux},
\author[hungary]{Z.~Fodor},
\author[clermont]{A.~Genoux-Lubain},
\author[hd]{G.~Goebels},
\author[strasbourg]{G.~Guillaume},
\author[hd]{E.~H\"afele},
\author[strasbourg]{F.~Jundt},
\author[hungary]{J.~Kecskemeti},
\author[moscow]{Y.~Korchagin},
\author[rossendorf]{R.~Kotte},
\author[strasbourg]{C.~Kuhn},
\author[moscow]{A.~Lebedev},
\author[kurchatov]{A.~Lebedev},
\author[romania]{I.~Legrand},
\author[strasbourg]{C.~Maazouzi},
\author[kurchatov]{V.~Manko},
\author[rossendorf]{J.~M\"osner},
\author[hd]{S.~Mohren},
\author[romania]{D.~Moisa},
\author[rossendorf]{W.~Neubert},
\author[hd]{D.~Pelte},
\author[romania]{M.~Petrovici},
\author[clermont]{P. Pras},
\author[strasbourg]{F.~Rami},
\author[strasbourg]{C.~Roy},
\author[hungary]{Z.~Seres},
\author[warsaw]{B.~Sikora},
\author[romania]{V.~Simion},
\author[warsaw]{K.~Siwek-Wilczy\'{n}ska},
\author[moscow]{V.~Smolyankin},
\author[moscow]{A.~Somov},
\author[strasbourg]{L.~Tizniti},
\author[hd]{M.~Trzaska},
\author[kurchatov]{M.~A.~Vasiliev},
\author[strasbourg]{P.~Wagner},
\author[rossendorf]{D.~Wohlfarth},
\author[kurchatov]{I.~Yushmanov}

\address[romania]{Institute for Nuclear Physics and Engineering, 
                  Bucharest, Romania}
\address[hungary]{Central Research Institute for Physics, Budapest, Hungary}
\address[clermont]{Laboratoire de Physique Corpusculaire, IN2P3/CNRS, 
                  and Universit\'{e} Blaise Pascal, Clermont-Ferrand, France}
\address[gsi]{Gesellschaft f\"ur Schwerionenforschung, Darmstadt, Germany}
\address[rossendorf]{Forschungszentrum Rossendorf, Dresden, Germany}
\address[hd]{Physikalisches Institut der Universit\"at Heidelberg, 
             Heidelberg, Germany}
\address[moscow]{Institute for Theoretical and Experimental Physics, 
                 Moscow, Russia}
\address[kurchatov]{Kurchatov Institute, Moscow, Russia}
\address[strasbourg]{Centre de Recherches Nucl\'{e}aires and 
               Universit\'{e} Louis Pasteur, Strasbourg, France}
\address[warsaw]{Institute of Experimental Physics, Warsaw University, Poland}
\address[rudjer]{Rudjer Boskovic Institute, Zagreb, Croatia}


\begin{abstract}
   Semi-inclusive triple differential multiplicity distributions of
positively charged kaons have been measured over a wide range in
rapidity and transverse mass for central collisions of $^{58}$Ni with
$^{58}$Ni nuclei. The transverse mass ($m_t$) spectra have been
studied as a function of rapidity at a beam energy 1.93~AGeV.  The
$m_t$ distributions of K$^+$ mesons are well described by a single
Boltzmann-type function.  The spectral slopes are similar to that of
the protons indicating that rescattering plays a significant role in
the propagation of the kaon.  Multiplicity densities have been
obtained as a function of rapidity by extrapolating the Boltzmann-type
fits to the measured distributions over the remaining phase space. The
total K$^+$ meson yield has been determined at beam energies of 1.06,
1.45, and 1.93~AGeV, and is presented in comparison to existing
data. The low total yield indicates that the $K^+$ meson can not be
explained within a hadro-chemical equilibrium scenario, therefore
indicating that the yield does remain sensitive to effects related to
its production processes such as the equation of state of nuclear
matter and/or modifications to the K$^+$ dispersion relation.
\end{abstract}
\end{frontmatter}

\section{INTRODUCTION}

  One of the main topics addressed in the study of relativistic
nuclear collisions is the question whether hadronic properties undergo
modifications in an environment of hot and dense nuclear matter. In
particular, theoretical predictions indicate that the self energy, or
effective mass, of hadrons may change in an environment of high
density and temperature~\cite{Ko97}.  Indeed, recent results collected
via leptonic probes at the SPS/CERN~\cite{CERES,HELIOS} have been
interpreted~\cite{dilep} as evidence that the mass of the $\rho$ meson
has dropped by a large fraction of its vacuum mass in the measured
system. To address this topic via the much more abundant hadronic
observables requires a detailed understanding of how the chosen probes
propagate through the dynamically expanding system after the moment of
highest density has been attained. In this context the behavior of
particles with strange quark content have attracted much theoretical
and experimental attention for some time~\cite{strangeness}. Due to
the small elastic cross-section ($\sim 10$~mb) and the conservation of
strangeness in the strong interaction, the yield and phase space
distributions of $K^+$ mesons have been considered a promising tool to
probe the dense matter formed in the initial stage of the
collision~\cite{kaonprod}. Furthermore, numerous calculations show
that K$^+$ meson production in nucleus-nucleus collisions is
intimately related to the mean potential or the self-energies of the
kaons in dense nuclear matter~\cite{pots}.  Such modifications to the
in-medium properties of K$^+$ mesons are also predicted to persist at
beam energies of 1 to 2~AGeV where baryonic densities of up to 3 times
normal nuclear matter is expected to be reached in the initial stages
of the reactions. At these beam energies which are near to or below
the K$^+$ meson production threshold in free nucleon-nucleon
collisions, kaon production in first chance nucleon-nucleon collisions
is suppressed.  As a result transport model calculations for nucleus
nucleus collisions indicate that the K$^+$ meson yield and spectra are
very sensitive to the nuclear equation of state
(EOS)~{\cite{EOS}}. Primarily this is because the yield is strongly
dependent on the rate of multi-step processes which increases rapidly
with increasing density. However, for lighter systems above threshold
it is expected that the EOS has little effect on the K$^+$ meson
yield~\cite{Huangpriv,noeos}. Instead, the dominating factor here is
the effective threshold for K$^+$ production in any individual
baryon-baryon collision, which is sensitive to the kaons in-medium
properties such as its dispersion relation.

  In an effort to collect evidence on whether changes to the K$^+$
meson's properties in a high baryon density region can be confirmed,
collisions of $^{58}$Ni with $^{58}$Ni nuclei have been measured near
the free nucleon-nucleon production threshold. After a short
description of the experimental device and the method of particle
identification, the K$^+$ meson differential cross sections are
presented as a function of transverse mass and rapidity, thereby
allowing a sensitive test of the extent to which the available phase
space is evenly populated. These data are compared to a similar
analysis of $\pi^-$ and proton distributions~\cite{Hong97a}. Next the
total K$^+$ meson multiplicity is estimated and presented as a
function of the beam energy. These results are compared to predictions
within a thermal model and discussed in the context of transport model
calculations. Finally, the results are summarized.

\section{EXPERIMENTAL SETUP}
\subsection{Detector}

  Collisions of the nuclei $^{58}$Ni on $^{58}$Ni have been measured
at incident energies of 1.06, 1.45, and 1.93~AGeV by the FOPI detector
at the SIS accelerator at GSI/Darmstadt.  The FOPI detector is a
highly modular system of several subdetectors for fixed target
experiments~{\cite{Gob93,Ritman95}}. It consists of 2 main parts, each
with nearly complete azimuthal symmetry. One main component is the
highly segmented Forward Wall (FW) which covers the polar angles
between $1^\circ < \Theta_{Lab} < 30^\circ$ over the full azimuthal
range. It provides the energy loss $\Delta E$ and position information
in addition to a Time--of--Flight (ToF) measurement, thus allowing
velocity and charge determination. This setup is supplemented by the
Central Drift Chamber (CDC), a cylindrical tracking detector covering
an additional range of polar angle from $32^\circ < \Theta_{Lab} <
150^\circ$.  The CDC is a drift chamber of the jet type which operates
at atmospheric pressure and is 200 cm long with inner and outer radii
of 20 and 80 cm respectively. In the azimuthal direction, the chamber
is divided into 16 sectors each containing 60 sense wires. For charged
particles, each sense wire provides an accurate measurement of the
radial and azimuthal (r and $\phi$) coordinates via the recorded drift
time ($\sigma_{r\phi} = 400$~$\mu$m) and a less precise measurement of
the z coordinate along the beam axis by charge division
$\sigma_z=15$~cm for minimum ionizing particles. The CDC operates
inside a magnetic field of 0.6~T produced by a superconducting
solenoid thus allowing the momentum of charged particles to be
determined. The transverse momentum resolution $\sigma $($p_t$)$/p_t$
has been determined by simulations with the GEANT package~\cite{GEANT}
to be about 4\% for $p_t < 0.5$~GeV (c$\equiv$1) and worsens to 9\%
near $p_t=1.5$~GeV~\cite{Pelte96}. The polar angle resolution is
better than $\sigma(\Theta) < 5^\circ$. Furthermore, the CDC provides
a multiple sampling of the specific energy loss $dE/dx$ with a
relative energy resolution $\sigma$($<dE/dx>$)/$<dE/dx>$ of about 15\%
for minimum ionizing particles.

\subsection{Particle Identification}

  Although negative pions can be unambiguously identified at all
momenta by the direction of curvature in the magnetic field ($K^-$
mesons are neglected here since their relative yield is on the
$10^{-4}$ level), the identification of positively charged particles
in the CDC relies on the Bethe--Bloch relation of the mean energy loss
$<dE/dx>$ of a track with its laboratory momentum $p_{Lab}$. As a
result, $\pi^+$ mesons can only be identified up to $p_{Lab}=650$~MeV
where the $<dE/dx>$ of the much more numerous protons approaches that
of the pions. In order to enhance the particle identification
capabilities of the CDC, the tracks are extrapolated to the
Scintillator and \v Cerenkov array (Barrel) that surrounds the CDC and
matched with the appropriate Barrel hit. The Barrel covers the polar
angular range of $44^\circ < \Theta_{Lab} < 160^\circ$ and has a time
of flight resolution of about $\sigma_{ToF}= 300$~ps over the full
length of the detectors. During the measurements reported on here only
one third of the total azimuthal range was covered by the Barrel. The
velocity is determined by dividing the path length as measured with
the CDC by the time of flight from the Barrel.

\begin{figure}[tht]
    \centering
    \mbox{\epsfig{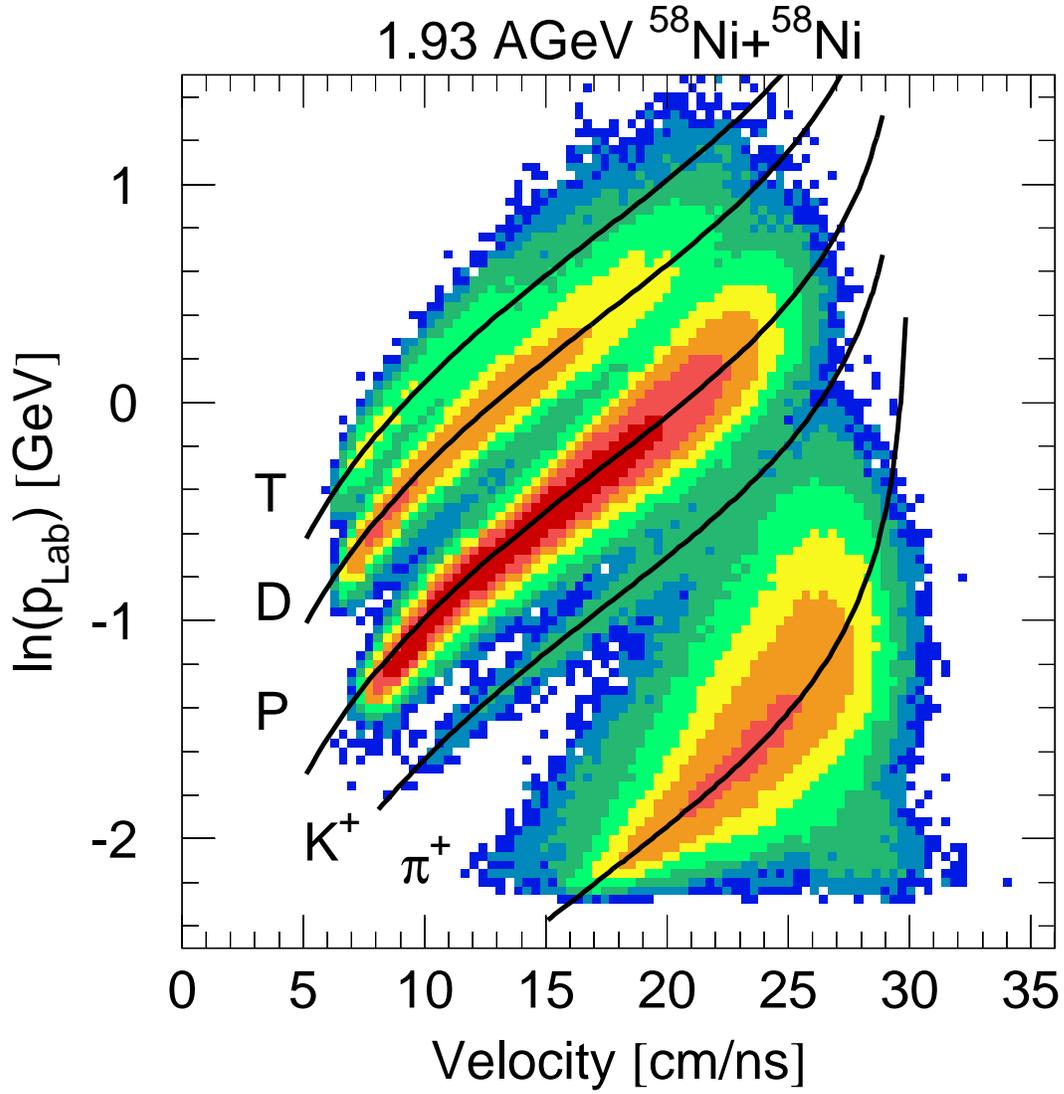}}
    \caption[]{Natural Logarithm of the laboratory momentum plotted as a 
     function of velocity for particles with charge=+1, presented with 
     contours of logarithmically increasing intensity.}
    \label{fig:p-v}
\end{figure}

\begin{figure}[tht]
  \centering
  \mbox{\epsfig{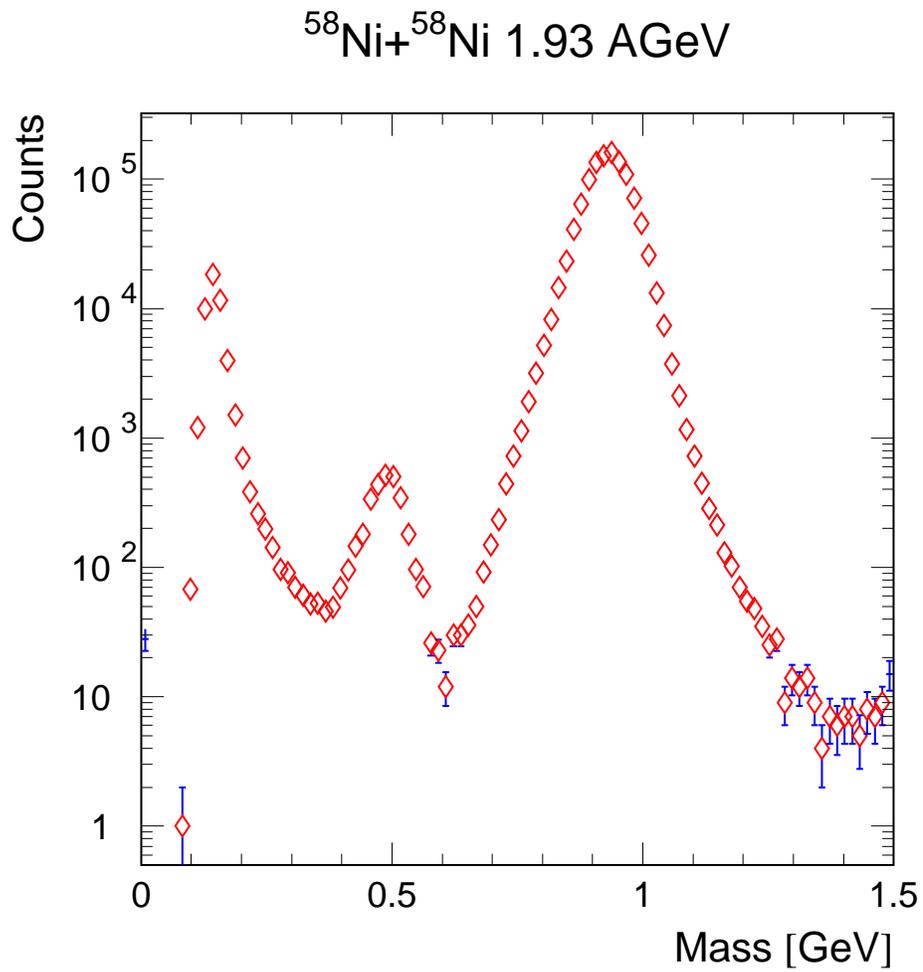}}
  \caption[]{Reconstructed mass of charge +1 particles with p$_{Lab} < $ 
    500~MeV.}
  \label{fig:k+k-}
\end{figure}

  Particles of different charge are distinguished by combining the
$<$~$dE/dx$~$>$ information with the velocity. After selecting charge
+1 particles, the natural logarithm of $p_{Lab}$ is shown plotted as a
function of the velocity in Figure~\ref{fig:p-v}. In this figure lines
corresponding to pions, protons, deuterons, and tritons are clearly
resolved over the full velocity range. In addition a line between the
pions and the protons is visible which results from the K$^+$
mesons. After an upper limit of $p_{Lab}<500$~MeV has been applied,
the mass spectrum is obtained as shown in Figure \ref{fig:k+k-}. A
peak at the kaon mass of 494~MeV is clearly visible above a background
arising from the $\pi^+$ and proton peaks.  A total of about 3000,
500, and 120 K$^+$ mesons have been identified within the
2.6$\times$10$^6$, 1.4$\times$10$^6$, and 1.5$\times$10$^6$ central
events collected at beam energies of 1.93, 1.45, and 1.06~AGeV,
respectively. Further details on the particle identification procedure
and the acceptance investigations can be found in
Ref.~{\cite{Ritman95,Bes96}}.

\begin{figure}[tht]
  \centering
  \mbox{\epsfig{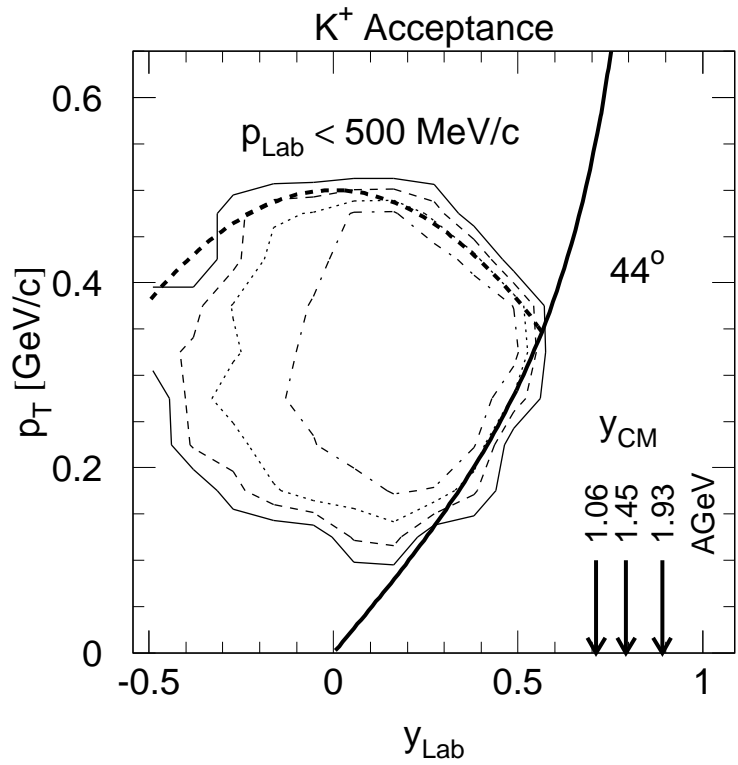}}
  \caption[]{Transverse momentum versus rapidity
    for K$^+$ mesons measured in Ni+Ni collisions at 1.93~AGeV. The solid line 
    denotes the detector boundary at 44$^\circ$, and the dashed curve 
    corresponds to the maximum p$_{Lab}$ for clean kaon identification. The 
    arrows denote mid-rapidity for the different beam energies studied here.}
  \label{fig:bar_pty_k+}
\end{figure}

 The strength of the FOPI detector is its ability to simultaneously
measure large regions of phase space. However, to insure clean K$^+$
meson identification, only kaons within the acceptance of the Barrel
sub-system have been analyzed.  To illustrate which region of phase
space has been analyzed, the transverse momentum ($p_t$) versus
rapidity ($y_{Lab}$) is plotted for K$^+$ mesons in
Figure~\ref{fig:bar_pty_k+}. As denoted by the solid line in this
figure, the acceptance of the Barrel starts at a laboratory polar
angle of 44$^\circ$. The upper momentum at which K$^+$ mesons can be
cleanly identified is about 500~MeV which is marked by the dashed
curve. The population at lower momenta is suppressed because the kaon
lifetime ($c\cdot\tau=371$~cm)~\cite{pp} is of the same order as the
typical flight path of about 125 cm. In addition, an absolute lower
transverse momentum limit of 100~MeV exists since the 0.6~T solenoidal
magnetic field prevents particles with lower $p_t$ from reaching the
Barrel.  For comparison, the arrows mark the location of mid-rapidity
at the beam energies studied here.

\subsection{Event Selection}\label{sec:apart}

  An ensemble of events with emphasis on small impact parameters (b)
has been collected. The event selection criterion exploited the strong
correlation between the impact parameter and the total particle
multiplicity (PMUL) observed in the reaction between $7^\circ <
\Theta_{Lab} < 30^\circ$.  The connection of PMUL to b is made by
taking the differential cross section $d\sigma/dPMUL$ to be a
monotonous function of $d\sigma/db$, whereby the observed cross
section has been determined from the integrated beam luminosity, the
target thickness (225 mg/cm$^2$), and the data acquisition deadtime.
The events analyzed here were taken from the 350 mb with the highest
PMUL, which would correspond to b less than 3.3 fm within a sharp
cutoff model. The actual ensemble however, includes events with
considerably higher b due to the large multiplicity fluctuations in
this small system. In order to facilitate the comparison of these
results with other experiments and theoretical predictions, an often
used convention is followed in which the average number of nucleons
with geometrical overlap ($A_{part}$) has been
calculated~\cite{Nifenecker85}.  After taking the b fluctuations into
account, the mean value $<A_{part}> \approx 75$ is estimated for the
central events investigated here.

\section{RESULTS}
\subsection{Transverse Mass Spectra}

 The differential multiplicity distribution $d^3N/dp^3$ of K$^+$
mesons has been measured in a wide range of rapidity and $p_t$ and
corrected for in-flight decay as well as known detector acceptance
effects. In order to facilitate comparisons of results collected at
different beam energies the scaled rapidity ($y^{(0)} = y/y_{CM}-1$) is
used, where target rapidity is at -1 and midrapidity is at 0. Since
these data show no significant azimuthal anisotropy~\cite{Rit95}, this
degree of freedom has been integrated and the results are presented as
a function of the transverse mass ($m_t = \sqrt{p_t^2 + m_{K^+}^2}$)
and rapidity. This representation has been chosen because a thermal
Boltzmann source would produce $m_t$ spectra of the following simple
form for a narrow window in rapidity $dy^{(0)}$:
\begin{equation}
	 \frac{1}{m_t^2}\frac{d^2N}{dm_tdy^{(0)}} =
	  I \cdot \exp\left[\frac{-(m_t-m_{K^+})}{T_B}\right]
\end{equation}
where both the Boltzmann inverse slope parameter ($T_B$) and the $p_t = 0$
intercept ($I$) are functions of rapidity. For a given isotropic thermal
source of temperature T, the $T_B$ values are not constant, but instead possess
the following functional dependence on rapidity:
\begin{equation}
	  T_B = T/cosh(y-y_{CM}).
\label{eq:cosh}
\end{equation}

\begin{figure}[tht]
  \centering
  \mbox{\epsfig{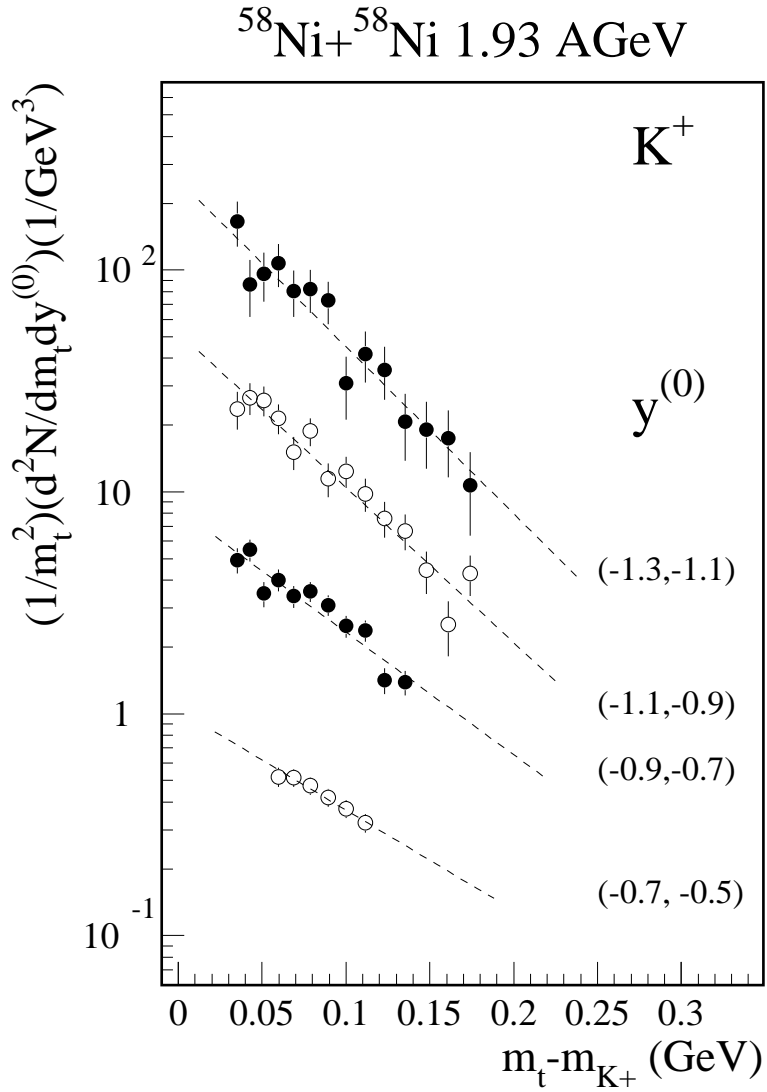}}
  \caption[]{Measured K$^+$ m$_t$ spectra for various slices in normalized 
    rapidity collected for the most central 350 mb of the system 
    $^{58}$Ni+$^{58}$Ni at 1.93~AGeV.  The spectra have been multiplied by 
    $10^n$ starting with n=0 for the range $-0.7<y^{(0)}<-0.5$.}
  \label{bare_mt2_k}
\end{figure}

  Figure~\ref{bare_mt2_k} shows the measured positive kaon spectra as a
function of ($m_t-m_{K^+}$) for several rapidity slices in units of
$GeV^{-3}$.  These data were collected for central events (see
Sec~\ref{sec:apart}) of $^{58}$Ni+$^{58}$Ni collisions at 1.93~AGeV.
The lowest spectrum corresponds to the highest rapidity value, and the
$n$th spectrum has been multiplied by 10$^n$ starting with $n=0$ at the
most central rapidity window. Within the detector acceptance and the
statistical fluctuations, the spectra are well described by the
exponential functions above. Each spectrum has been fit with this
function and the results are represented by the dashed lines in
Figure~\ref{bare_mt2_k}. The overall distribution of $\chi^2$ per
degree of freedom is between 0.5--2; however, estimates of the total
yield (in the following section) must take into account the negative
correlation between the two parameters $I$ and $T_B$. Since this
procedure is carried out at many narrow rapidity intervals, both $T_B$
and $I$ have been determined as functions of rapidity, and the $T_B$
values are listed below in Table~\ref{tab:data}.

\begin{figure}[tht]
  \centering
  \mbox{\epsfig{file=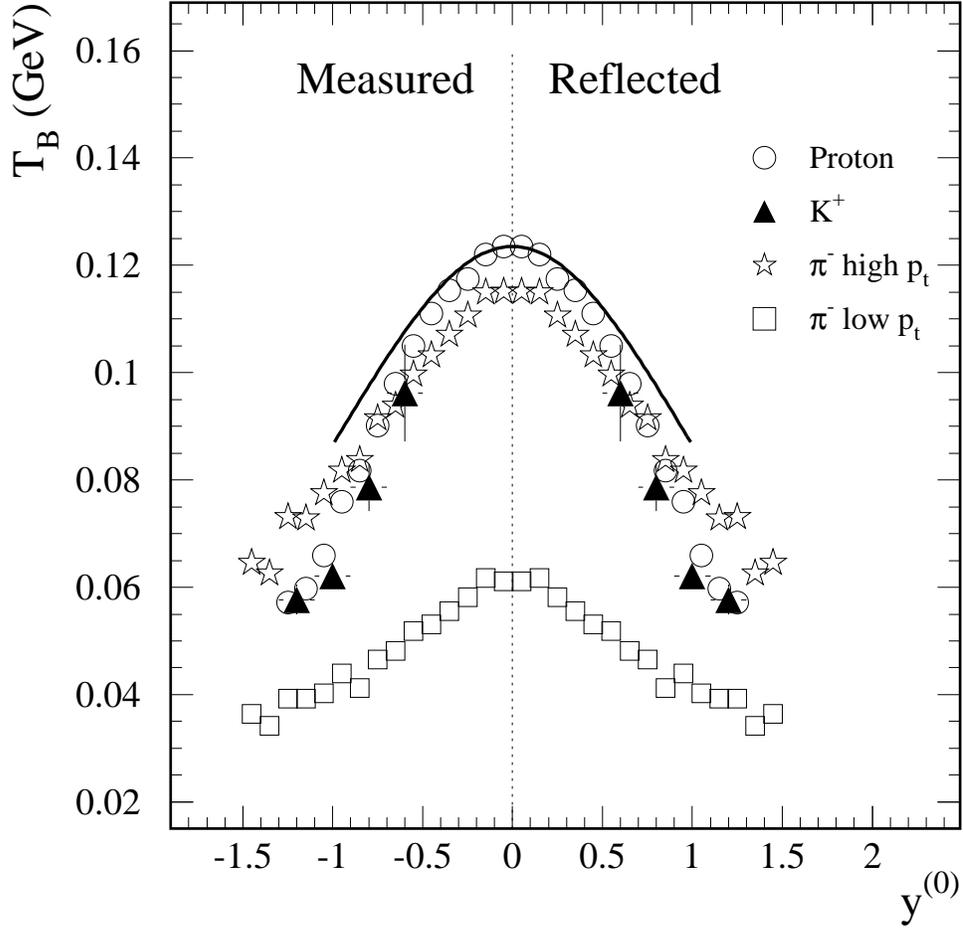}}
  \caption{Measured slope parameters of $\pi^-$, K$^+$, and protons as a 
    function of normalized rapidity for the system $^{58}$Ni+$^{58}$Ni at 
    1.93~AGeV. The solid line marks the expected behavior of a thermal source
    with a temperature given by the proton's mid-rapidity slope.}
  \label{all_tb_y_all}
\end{figure}

 The proton and $\pi^-$ spectra have also been
studied~\cite{Hong97a}. While the protons can be well described by a
single exponential in all rapidity intervals, the pion spectra exhibit
a concave shape, which can be described by the sum of two such
exponential functions with inverse slope parameters $T_B^L$ and
$T_B^H$ for the softer and stiffer components, respectively. The
inverse slope parameters of the K$^+$ mesons, protons, and both pion
components are plotted together in Figure~\ref{all_tb_y_all} for the
different rapidity slices. The symbols of the left indicate the
measured data which have been reflected onto the right side for
symmetry reasons. At midrapidity the proton slopes are higher than the
pion. This observation may indicate that the system contains a
collective expansion velocity in addition to a thermal component. The
magnitude of the K$^+$ meson slopes near target rapidity is similar to
or lower than the proton slopes, which is qualitatively in agreement
with the midrapidity results from the KaoS
collaboration~\cite{Senger96}, but in contrast to the somewhat harder
K$^+$ than proton spectra as obtained by Schnetzer et al. in the
system Ne+NaF at 2.1~AGeV~{\cite{schnetzer}}.

 Away from midrapidity ($y^{(0)}=0$) the measured inverse slope
parameters $T_B$ of all particles decrease with increasing
$|y^{(0)}|$. For the pions, the rapidity distribution of the $T_B^H$
values follow the function in Eq.~\ref{eq:cosh} for $|y^{(0)}| < 1$,
in agreement with isotropic emission from a source at
midrapidity. This shape however is rather insensitive to the presence
or absence of radial flow due to the small pion mass.  The pion's
$T_B^L$ distribution is somewhat flatter than expected from
Eq.~\ref{eq:cosh}, which is in agreement with suggestions that this
component is governed primarily by the decay kinematics of the
$\Delta$ resonance~\cite{delta}. On the other hand, the proton $T_B$
distribution (open circles) is noticeably narrower than the
$T/cosh(y)$ dependence which is shown by the solid line in
Fig.~\ref{all_tb_y_all}, thus indicating a more complex source
behavior. The $T_B$ distribution for the K$^+$ mesons has a form
similar to that of the protons indicating significant elastic K$^+$N
scattering.

\subsection{Rapidity Distributions}

In order to determine the multiplicity density $dN/dy^{(0)}$, the
exponential fits to the double differential yields are integrated from
$m_t = m_{K^+}$ to $\infty$. This integral can be expressed in the
following closed form:
\begin{equation}
	 \frac{dN}{dy^{(0)}} \propto 
	  I \cdot exp\left(-\frac{1}{k}\right) \cdot 
	  \left[2+\frac{2}{k}+\frac{1}{k^2}\right] \cdot T_B^3 
\end{equation}
with $k=T_B$/m$_{K^+}$. The statistical error for $dN/dy^{(0)}$
(standard deviation) is evaluated from the uncertainty of the right
term of this formula using the variances and covariances of $I$ and
$T_B$. The $dN/dy^{(0)}$ values for $K^+$ mesons which have been
determined by this procedure are presented in Table~\ref{tab:data}
together with the $T_B$ parameters. The values quoted include the
statistical and systematic uncertainties (see Sec.~\ref{sec:errors}),
respectively.
\begin{table}[hhh] 
\begin{center}
\begin{tabular}{cccc}
\hline \hline
$y^{(0)}$   &  $T_B$ (MeV)        & $T=T_B/\cosh(y)$& $dN/dy^{(0)}\times10^{3}$\\
-1.2	    & $57.7\pm2.8\pm5.8$  &$94.0\pm4.5\pm9.4$ & $ 4.5\pm0.3\pm0.6$ \\
-1.0        & $62.1\pm2.5\pm6.2$  &$88.2\pm3.5\pm8.8$ & $10.2\pm0.4\pm1.3$ \\
-0.8        & $78.6\pm4.4\pm7.9$  &$99.5\pm5.6\pm10.0$& $21.8\pm1.2\pm3.2$ \\
-0.6        & $96.2\pm9.0\pm9.6$  &$109.7\pm10.3\pm11.0$& $36.0\pm2.4\pm7.2$ \\
\hline \hline
\end{tabular}
\end{center}
\caption[]{Measured K$^+$ inverse slope parameters, apparent temperatures, and
multiplicity densities collected for central events of the system
$^{58}$Ni+$^{58}$Ni at 1.93~AGeV. The listed uncertainties are statistical and
systematic, respectively.}
\label{tab:data}
\end{table}

\begin{figure}[tht]
  \centering
  \mbox{\epsfig{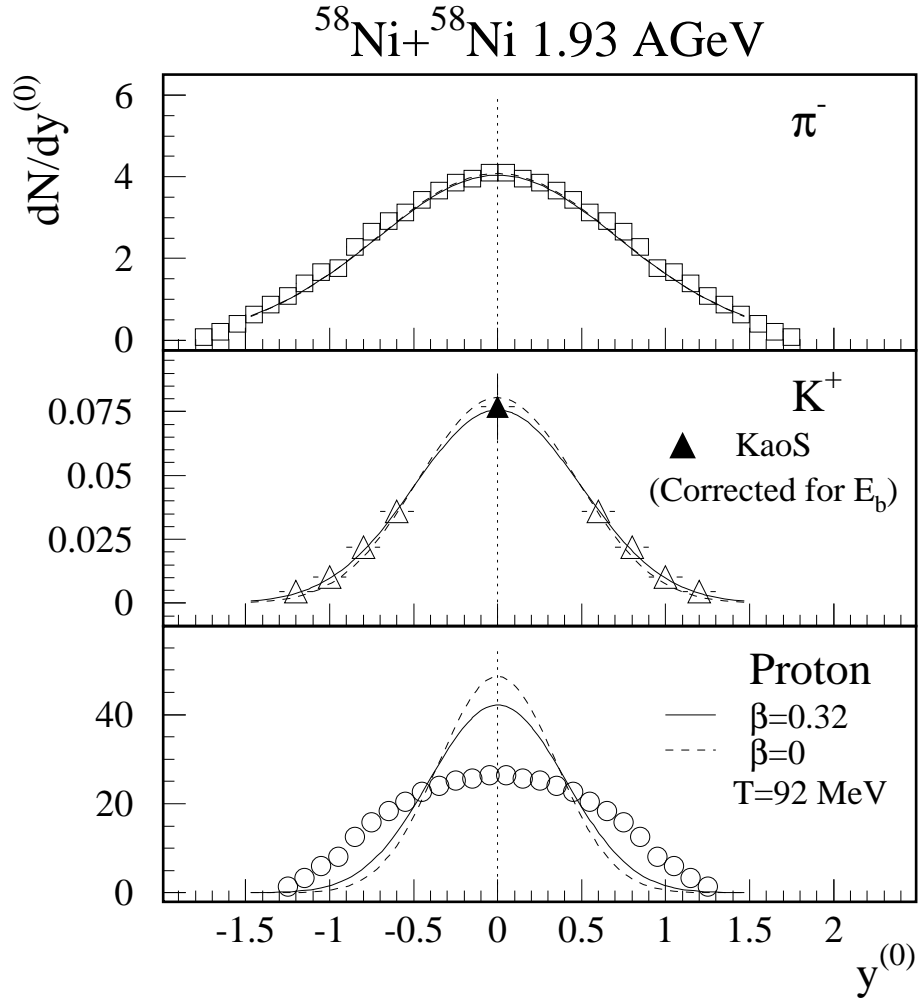}}
  \caption{Deduced multiplicity density dN/dy$^{(0)}$ of pions, kaons, and
    protons for the system $^{58}$Ni+$^{58}$Ni at 1.93~AGeV. The data points 
    to the right of the dashed line are reflections of the points on the left.}
  \label{all_dn_dy_all}
\end{figure}

  The rapidity distributions of positive kaons are shown in
Figure~\ref{all_dn_dy_all} in comparison to the $\pi^-$ and proton results.
Again the measured points on the left side have been reflected around
midrapidity. The dashed lines represent the expected distribution for a purely
thermal source located at midrapidity, and the solid lines shown in this
figure represent the expected distribution for a radially expanding thermal
source~\cite{Siemens79}, which has been shown to be an essential component to
describe the energy and rapidity spectra (see e.g.~\cite{Jeong,Lisa}). In this
case the temperature (T=92~MeV) and flow velocity ($\beta$=0.32) parameters
were determined by a simultaneous fit to the $\pi^-$, proton, and deuteron
midrapidity spectra~\cite{Hong97a} and the yield was normalized to the data.

 The $dN/dy^{(0)}$ distributions for the produced particles ($\pi^-$
and K$^+$) are well reproduced by the two curves showing a thermalized
source both with and without a radial flow component. In contrast, the
non-thermal behavior of the protons is very noticeable.  Although the
increased width of the proton rapidity distribution may be reproduced
by allowing for either significant transparency or longitudinal flow,
the increase is most probably a result of so called ``Halo"
effects which arise from the coarse centrality selection available
in this rather small system (Ni+Ni) due to the large fluctuations of
the observed particle multiplicity.

\subsection{Total Production Probability}

  The total production probability of pions and protons over the full
phase space has been determined at all three beam energies by
integrating the respective $dN/dy^{(0)}$ spectra.  However, for K$^+$
mesons this direct method was not possible at the lower beam energies
because the recorded statistics were insufficient to perform the above
mentioned differential analysis to obtain $dN/dy^{(0)}$, and at the
highest energy (1.93~AGeV) because mid-rapidity was not covered.
Therefore, to determine the total multiplicity, the form of the K$^+$
phase space distribution has been taken from an isotropic, radially
expanding source with the temperature and flow velocity parameters
determined from the mid-rapidity pion, proton, and deuteron spectra at
each beam energy as quoted above.  With decreasing beam energy, this
resulted in a detection efficiency that increased from about 5\% to
10\% of the total yield in the backward hemisphere, without including
decay losses. In support of this assumption of the source shape we
have added the midrapidity data from the KaoS collaboration to
Figure~\ref{all_dn_dy_all} as the filled triangle~\cite{Senger96}.
Since the KaoS measurement was taken at 1.8~AGeV, their result for an
equivalent centrality selection was increased by 45\% using the
following scaling:
\begin{equation}
   \sigma^{K^+} \propto E_{b}^\alpha
\label{eq:kbeam}
\end{equation}
where the exponent of the beam energy $E_b$ was determined to be
$\alpha = {5.3\pm0.3}$ from the K$^+$ excitation function discussed below. 

  The total K$^+$ meson multiplicity in the central events measured
here are listed in Table~\ref{tab:kprod}. In order to facilitate the
comparison of these exclusive results with other experimental data,
the yield is expressed relative to the mean number of participating
nucleons $<A_{part}>$.
\begin{table}[hhh] 
\begin{center}
\begin{tabular}{cc}
\hline
\hline
 $E_{b}$    &  $P^{K^+}_{NN}$                \\
 1.060 AGeV &  (4.4$\pm$1.1) $\times 10^{-5}$\\
 1.450 AGeV &  (2.6$\pm$0.65)$\times 10^{-4}$\\
 1.930 AGeV &  (1.1$\pm$0.28)$\times 10^{-3}$\\
\hline
\hline
\end{tabular}
\end{center}
\caption[]{K$^+$ production probability per participating nucleon at the
various beam energies measured.}
\label{tab:kprod}
\end{table}
The data in Table~\ref{tab:kprod} have thus been scaled by a factor
$<A_{part}> = 75$ (see Section~\ref{sec:apart}). The errors quoted in this
table are dominated by the systematic uncertainty in the extrapolation over
the full phase space.

\begin{figure}[tht]
  \centering
  \mbox{\epsfig{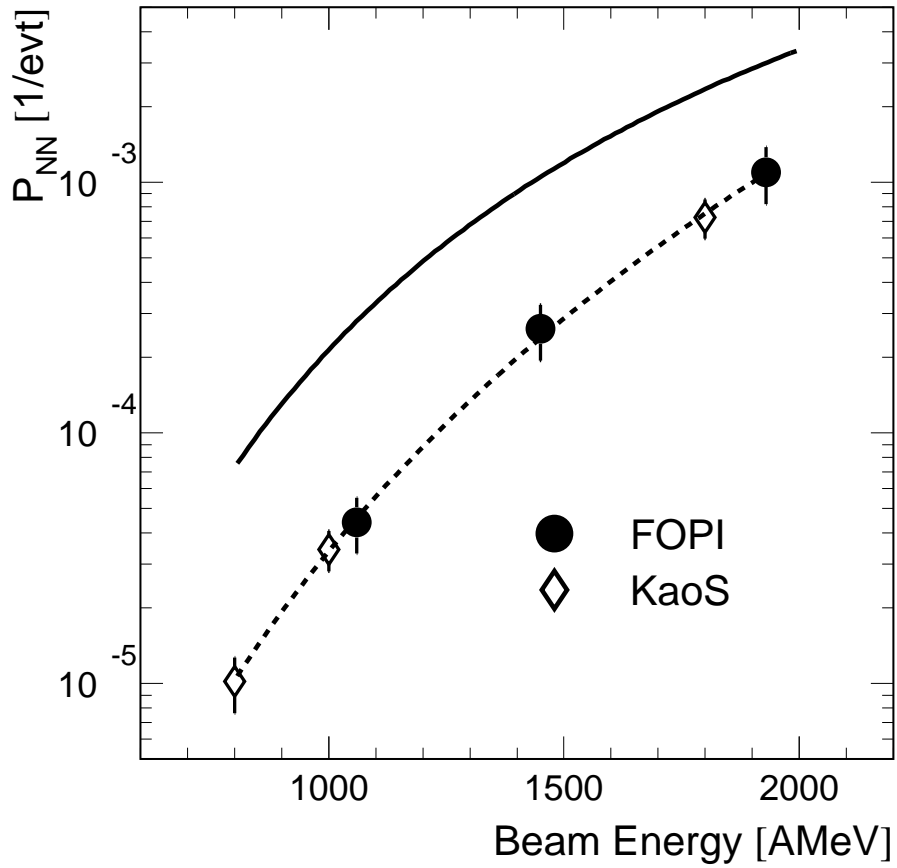}}
  \caption{Kaon production probability per participating nucleon as a function 
    of beam energy for the system Ni+Ni. The curves are described in the text.}
  \label{kaon_sys}
\end{figure}

  For comparison, these data values have been plotted in
Figure~\ref{kaon_sys} together with the Ni + Ni results from the KaoS
collaboration~\cite{Senger96}.  The KaoS results which were obtained
in a narrow window at mid-rapidity have been scaled by the reaction
cross section (2.71 b) and by the mean number of participating
nucleons, which is $A_{proj}/2$ for an inclusive measurement of a
symmetric system within a geometric model~\cite{Nifenecker85}.
Although this normalization of the yield by $<A_{part}>$ is quite
successful for pions in small and medium mass systems~\cite{Harris87},
for large systems (Au) at sub-threshold energies~\cite{Miskowiec} this
ratio does vary with centrality for K$^+$ mesons. The dashed line in
this figure shows a fit with the function presented in
Equation~\ref{eq:kbeam} which results in an exponent
$\alpha={5.3\pm0.3}$. The solid line marks the expected production
rate from the empirical scaling presented by V.~Metag~\cite{Metag93}
after multiplying the abscissa by the K$^+$ free production threshold.
This curve essentially shows the pion production rate that would be
expected if it had the same free production threshold as the K$^+$ meson.
The data clearly lie below this curve; however with increasing beam
energy (from 1 to 2~AGeV) the magnitude of the discrepancy is reduced by
a factor two. This large deviation of the K$^+$ yield from the scaling
is a first indication that the kaons do not go through a state of
chemical equilibrium at these beam energies.

\subsection{Systematic Errors}\label{sec:errors}

  The uncertainties shown in figures 4-6 are statistical errors, and
errors smaller than the symbol size are not shown. The systematic
error on the inverse slope parameters $T_B$ is about 10\%, which is
largely due to the rapidity binning and the range of $m_t - m_{K^+}$
used in the fitting procedure, as well as uncertainties arising from
the poor polar angle resolution. The systematic errors for the
$dN/dy^{(0)}$ results vary with rapidity and are estimated to vary
from 13\% to 20\%. This comes from errors in tracking efficiencies
(10\%), particle identification (2\%), and from the extrapolation
procedure (from 8\% to 15\% with increasing rapidity).  These errors
are assumed to be incoherent. Tracking efficiency was obtained both by
comparing the results of visual scanning of several hundred events
with the output of the tracking program~\cite{Bes96}, and by comparing
the results of several different tracking algorithms based on local
and global tracking methods respectively~\cite{Pelte96}. The errors
for particle identification are estimated from variations made in the
software selections.  The systematic uncertainty arising from the
extrapolation procedure of the measured $m_t$ spectra over the full
$m_t$ range was estimated by comparing the total yield of all charged
reaction products (up to $^4$He) to the initial $^{58}$Ni nuclei.
Although only 57\% of the total charge was actually measured, after
the extrapolation procedure, the total charge was reproduced to within
5\%.

  The background estimation for the K$^+$ meson spectra has been
performed as follows: The mass spectra have been investigated in the
measured laboratory momentum range of 100 to 500~MeV with a binning of
20~MeV. Since the structure of the background is not precisely known,
a worst case analysis has been applied.  Since the form of the mass
spectrum (see Fig.~\ref{fig:k+k-}) between any particle pair is
concave, an upper limit of the background contribution is estimated by
connecting the minimum between the proton and the K$^+$ meson peaks
with the minimum between the K$^+$ and the $\pi^+$ peaks using an
exponential function.  The integration of this exponential over the
range $\pm 2\sigma$ from the K$^+$ peak gives an upper limit of the
background contribution to the kaon yield. This contribution is
negligible below 400~MeV then rises sharply to reach 20\% at 500~MeV.
In comparison, the correction for the in-flight decay of the K$^+$ is
assumed to introduce little systematic error since the lifetime is
very well known and the measured fraction increases from 45\% at a
laboratory momentum of 200 MeV to over 70\% at 500~MeV.

\section{DISCUSSION}
 
  This section begins with an attempt to determine the extent to which
the K$^+$ mesons deviate from uniform phase space population. This is
important because a system that goes through a phase of equilibrium
loses much information about the specific production processes prior
to when the particle has decoupled from its surrounding
environment. Following this, the results from several theoretical
models are compared to these measurements in an attempt to determine
whether any modification of the K$^+$ meson's properties during the
dense stage of the reaction can be inferred from these data.

\subsection{Approach to Equilibrium}

  To address the question to what extent at least a part of the system
is in thermal equilibrium, the following observations on the phase
space distributions can be drawn from the data presented here: \\
  Globally the nucleons display large deviations from isotropic thermal
behavior. However, the proton's elongated $dN/dy^{(0)}$ distribution
by itself is not sufficient to differentiate whether the nucleons can
be described in terms of a single component that is only partially
stopped (or fully stopped followed by longitudinal expansion) in the
collision, or if it is more appropriate to consider a two component
description where a sub-set of the initial system goes through a state
approaching thermal equilibrium, and the remainder proceeds relatively
unaffected by the collision. However, the high $T_B$ for protons at
midrapidity and the rapid drop with increasing normalized rapidity favor
the two-component description since transparency or longitudinal
expansion would rather lead to a wider $T_B$ distribution. \\ 
  Taking the softer component of the pions to be primarily a result of
the $\Delta$ decay kinematics, the remaining pions (hard component)
near mid-rapidity have $T_B$ and $dN/dY^{(0)}$ distributions that are
consistent with a state close to thermal equilibrium. \\
  The K$^+$ mesons have a similar $T_B$ distribution to the protons
indicating that rescattering plays a significant role, and as a result
the momentum distribution may remain sensitive to the propagation of
the kaons in the dense medium as discussed below. \\
  These observations are thus consistent with a picture where a portion
of the initial nucleons are stopped at mid-rapidity, building a
roughly thermalized system which is surrounded by the remaining cooler
nucleons which remain close to beam rapidity.

  The next step is to determine whether the measured particle yields
are consistent with expectations for a mid-rapidity source which is
also in chemical equilibrium.  Taking the formalism used in
reference~\cite{PBM}, the particle ratios such as K$^+$/nucleon can be
calculated including finite size effects and directly compared to the
measured values in Table~\ref{tab:kprod}.  The necessary parameters
have been determined as follows: the degeneracies from spin and
isospin are $g_N = 4$ for the nucleons and $g_{K^+}=1$ for the kaons
(only K$^+$ are under consideration). The temperature (92~MeV) and the
Baryon chemical potential ($\mu_B=0.7$~GeV) have been extracted from
the measured data~\cite{Hong97a,Hong96}.  The strangeness chemical
potential $\mu_S$ can be calculated within the model by itself (for a
given $T$ and $\mu_B$) by requiring strangeness conservation, i.e. the
total density of particles with $s$ quarks must equal the density of
particles with $\overline{s}$ quarks (i.e.  $\rho_{K} =
\rho_{\overline{K}}+\rho_{\Lambda}+\rho_{\Sigma}$).  For this $T$ and
$\mu_B$, a value $\mu_S=0.135$~GeV results~\footnote{ Alternatively,
$\mu_S$ can be deduced by the ratio K$^-$/K$^+=\exp
(-2\mu_S/T)$. Taking the measured ratio K$^-$/K$^+$ to be
0.03~\cite{k-k+ratio}, then $\mu_S=0.16$~GeV.}.  With this set of
parameters the measured pion yield can be well reproduced.
Furthermore, the production of the $\eta$ meson, which has a similar
free production threshold as the K$^+$ meson, has also been measured
in the same system ($^{58}$Ni + $^{58}$Ni at 1.9~AGeV)~\cite{App97}.
Here too the relative yield of the $\eta$ can be understood in terms
of chemical equilibrium.  In contrast to the non-strange mesons
however, the corresponding ratio K$^+$/nucleon = 3\% is in strong
disagreement with the measured result of 0.11\% at 1.93~AGeV.
Variation of these parameters within reasonable boundaries such that
the spectra remain well described can reduce the expected
K$^+$/nucleon ratio by at most a factor of 2, which still leaves an
order of magnitude discrepancy to the data.

  This interpretation of the positive kaon yield in terms of the
approach toward an equilibrium value is also reflected in the kaon
excitation function. The suppression of the K$^+$ yield below that of
the empirical scaling shown in Figure~\ref{kaon_sys} is observed to
decrease by more than a factor of two with increasing beam energy over
the range covered here. This convergence is consistent with
qualitative expectations that with increasing beam energy the K$^+$
meson source increases in temperature and density, thus reducing the
relaxation time $\tau_{K^+}$ and allowing the source to come closer to
its equilibrium value. Similarly, this interpretation might provide an
alternative explanation for the observed rise of the K$^+/A_{Part}$
ratio with source size as observed at both SIS and AGS
energies~\cite{Senger96,Wang96} without having to introduce a change
of the in-medium kaon mass~\cite{Ko95,Ehehalt96}.

  To summarize this discussion the following conclusions can be drawn:
the momentum distribution of K$^+$ mesons seems to have a large
influence from elastic scattering with the surrounding nucleons, and
the total K$^+$ yield is irreconcilable with a state that is also in
chemical equilibrium.

\subsection{Microscopic Model Calculations}

  These observations provide the basis from which details of the K$^+$
meson production processes might be extracted. However, due to the
complex nature of the system under discussion, it appears necessary to
compare the experimental results with microscopic transport model
calculations in order to separate effects arising from the behavior of
kaons in the initial, compressed state from the dynamics of the later
expanding stages. Several of the models that exist have been used to
calculate K$^+$ meson production in the system $^{58}$Ni+$^{58}$Ni at
1.93~AGeV. For instance, the IQMD model~{\cite{Hartnack}} predicts
that the kaons are produced with relatively small momenta (mainly due
to the limited phase space) and then on average gain energy by elastic
scattering with the surrounding baryons thereby hardening the K$^+$
spectra and widening the $dN/dy^{(0)}$ distribution. In fact, RBUU
calculations of this system with an impact parameter b=2~fm suggest
that a K$^+$ suffers on average over 2 elastic $K^+N$ scatterings
after its production~{\cite{Lipriv}}. Thus, the similarity between the
measured K$^+$ and proton slopes shown in Figure~\ref{all_tb_y_all}
seems to indicate a considerable amount of elastic scattering of K$^+$
mesons with the nucleons even in this relatively small system. Despite
this rescattering, the K$^+$ mesons should remain sensitive to the
dense stage of the reactions since within the QMD studies more than
60\% of the positive kaons have their last interaction at a density
greater than normal nuclear matter density. In order to make a more
detailed comparison of the data with these various model predictions,
the theoretical calculations must be subject to the appropriate
experimental filter in order to account for more subtle aspects of the
detector's acceptance and efficiency.

  The total K$^+$ meson yield has also been studied. However it has
been predicted that the sensitivity to the EOS is greatly reduced in
this rather small system size at the highest energy presented
here~\cite{Huangpriv}. Instead, a more sensitive measure of the EOS
might be obtained by trying to reproduce the shape of the kaon
excitation function. In fact it has been suggested that the
sub-threshold K$^+$ meson yield is sensitive not only to the bulk
compressibility of nuclear matter, but also to details in the shape of
the EOS, for instance the existence of density
isomers~\cite{Hartnack94}.  Within these microscopic model
calculations a jump in the excitation function by up to an order of
magnitude might be expected. However, the K$^+$ meson excitation
function shown in Fig.~\ref{kaon_sys} exhibits a smooth increase by
over two orders of magnitude as the beam energy is varied from deep
sub-threshold to well above the free production threshold. Since no
such discontinuity is seen in the data, the existence of density
isomers can not be confirmed.

  At this point a large amount of freedom still remains in a number of
input parameters to the theoretical calculations, for example the
cross sections of various meson-baryon and baryon-baryon
collisions. In order to overcome this uncertainty in the predictions
of the K$^+$ meson yield, it has been suggested that other aspects of
the phase space distribution may also retain information on the kaon
properties during the dense stage of the reaction for instance the
K$^+$ sideward flow may be sensitive to the kaon potential in the
dense nuclear medium~\cite{Li95}. In fact, calculations~\cite{gqli96a}
that have been compared to our data~\cite{Rit95} would imply that the
K$^+$ mesons do indeed feel a only weakly repulsive potential,
implying that there is a very attractive scalar component in this
dense nuclear medium. Other approaches however are able to reproduce
the measured flow values without having to introduce
potentials~\cite{David97}.

\section{SUMMARY}

  In this paper the production of positive kaons in central collisions
of $^{58}$Ni+$^{58}$Ni with beam energies from 1 to 2~AGeV has been
presented.  At the highest energy the $m_t$ spectra of the K$^+$
mesons can be described by single component Boltzmann functions. The
inverse slope parameters are similar to the protons and the
$dN/dy^{(0)}$ spectra are compatible with the hypothesis of a thermal
source with temperature 92 MeV and radial expansion velocity $\beta =
0.32$ which describes the pion, proton, and deuteron mid-rapidity
spectra. The K$^+$ meson excitation function rises smoothly, in
agreement with other measurements of Ni + Ni. The magnitude of the
yield however is far below that expected for a system in chemical
equilibrium. This non-equilibrium behavior is an important
prerequisite to extract details about the production processes from
the particle spectra in the final state. In an attempt to distinguish
production from propagation effects, transport models have been
employed.  Despite the expected sensitivity of some observables
(e.g. sub-threshold kaon yield to the EOS), it has been seen that
other effects can influence these observables by a similar
magnitude. Thus an analysis of the sideward flow could be very useful
to determine the magnitude of the mean potential felt by the kaons in
the dense nuclear medium. Although it is not yet possible to
unambiguously conclude whether the kaon mass changes in a system of
high density, an even greater sensitivity is expected with the
negative kaon results~\cite{Li9609}.

{\bf {\it {ACKNOWLEDGMENTS}}}

 We would like to dedicate this work to the memory of V.~L.~Krylov who played a
central role in the design and construction of the BARREL detector.

Furthermore we gratefully acknowledge the many useful discussions with 
G.~E.~Brown, C.~Hartnack, S.~W.~Huang, G.~Q.~Li, and P.~Senger, as well 
as the important comments from P.~Braun-Munzinger.

This work has been supported in part by the German Bundesministerium
f\"ur Forschung und Technologie under contracts 06 HD 525 I(3),
X051.25/RUM-005-95 and X081.25/N-119-95 and by the Polish State Committee of
Scientific Research (KBN) under grants 2 P302 1104 and 2 P03B 019 11.

\newpage

\end{document}